\documentclass[journal,9pt]{IEEEtran}
\usepackage{amsmath,amsfonts}
\usepackage{algorithmic}
\usepackage{algorithm}
\usepackage{array}
\usepackage{textcomp}
\usepackage{stfloats}
\usepackage{url}
\usepackage{verbatim}
\usepackage{graphicx}
\usepackage{cite}
\usepackage{tikz}
\usepackage{subcaption}
\usepackage{bm}  
\usepackage{multirow}

\usepackage{changes}

\usetikzlibrary{decorations.pathmorphing}
\usetikzlibrary{arrows,calc,decorations.markings}
\usetikzlibrary{arrows,shapes,calc,math}

\begin{document}

\title{\huge{Multilayer Dual-polarized Microstrip Antenna Design by\\ Topology Optimization with Enhanced Bandwidth}}

\author{\Large{Pan Lu, Eddie Wadbro, Viktor Lundström, Jonas Starck, Martin Berggren, Emadeldeen Hassan}

\thanks{This work was supported by the Swedish strategic research program eSSENCE and Kempestiftelserna. }
\thanks{P.\,Lu and M.\,Berggren are with the Department of Computing Science, Umeå University, 901\,87 Umeå, Sweden (e-mail: plu@cs.umu.se; martin.berggren@cs.umu.se). }
\thanks{E.\,Wadbro is with the Department of Mathematics and Computer Science, Karlstad University, 651\,88 Karlstad, Sweden, and the Department of Computing Science, Umeå University, 901\,87 Umeå, Sweden (e-mail: eddie.wadbro@kau.se).}
\thanks{J. Starck and Viktor Lundström are with Proant AB, 906\,20 Umeå, Sweden, and Mobile Mark Inc.(e-mail: jstarck@mobilemark.com; vlundstrom@mobilemark.com).}
\thanks{E.\,Hassan is with the Department of Applied Physics and Electronics, Umeå University,  901\,87 Umeå, Sweden (e-mail: emadeldeen.hassan@umu.se).}
}

\markboth{}%
{}


\maketitle

\begin{abstract}
Dual-polarized (DP) microstrip antennas are utilized in wireless systems for efficient data transmission.
However, their bandwidth is typically very limited.
In this contribution, we propose to design DP microstrip antennas with enhanced bandwidth using a density-based topology optimization approach. 
We formulate an optimization problem that simultaneously accounts for feeding port matching, the ports' isolation, and far-field dual-polarized performance. 
To enhance the bandwidth, we employ an FR4 stack-up, in which the copper on two layers is optimized simultaneously. 
We present two antenna designs operating around 5.7\,GHz, which show a compromise in performance between a high isolation (more than 40 dB) and enhanced impedance bandwidth (around 10\%). 
The optimized designs are experimentally validated, showing an excellent agreement between the simulated and measured performance.
\end{abstract}


\section{Introduction}
\IEEEPARstart{M}{icrostrip} antennas are widely used in modern wireless communication systems due to their low profile, low manufacturing cost, and easy integration with PCB circuits. 
To enhance channel capacity and improve signal diversity, wireless systems employ DP antennas\,\cite{mishrareview,10700661}.
Conventional microstrip antennas typically exhibit  a narrow bandwidth\,\cite{feng2019dual}, which limits their performance in modern systems such as satellite communications, MIMO, and radar systems. 
Therefore, the design of DP\,microstrip antennas with enhanced bandwidth is of great significance.

Traditional approaches to designing DP antennas rely heavily on theoretical analysis and the expertise of design engineers.
A common strategy to achieve dual polarizations is to impose structural symmetry on the design\,\cite{chen2023single,zhou2026ultra,zhu2022hybrid,he2020dual,wu2023wideband}.
Chen\,\cite{chen2023single} proposed a single-layer single-patch DP microstrip antenna based on higher-order mode analysis; however, only $1.8$\% impedance bandwidth was achieved.
To enhance the bandwidth, the use of multiple substrates has been suggested, such as in the work by Zhou et al.\,\cite{zhou2026ultra}, where an ultrawideband antenna is designed for 5G/6G systems based on a magneto--electric dipole. 
Li et al.\,\cite{li2023low} employed characteristic mode theory to design a wideband DP antenna for vehicular applications.

Optimization algorithms and neural network-based optimization offer systematic and effective approaches to design antennas with complex objectives\,\cite{Slawomir22Tolerance,li2023nn}.
In the last decade, density-based topology optimization (TO) has been successfully utilized to design various electromagnetic devices, including antennas\,\cite{zhu2022hybrid,wang2023efficient}, microwave\,\cite{ludecouple,bokhari2023topology,emad2020waveguide} and optical components\,\cite{Gedeon25Time}.
The large number of degrees of freedom offered by this approach enables designs that go far beyond human intuition, while also allowing manufacturing constraints to be incorporated into the optimization process using filtering techniques\,\cite{hassan2018topology}.  
Zhu et al\,\cite{zhu2022hybrid} proposed a hybrid TO method, combining density-based and level-set methods, to optimize a DP antenna, where metallic vias surrounding the structure are employed to improve the antenna beamwidth.

In this work, we propose a density-based topology optimization approach for the design of planar multilayer DP microstrip antennas with enhanced bandwidth.
Unlike conventional approaches that rely on predefined radiating element structures, the proposed method determines the optimal conductive distribution in a fully algorithm-driven manner. 
The optimization problem is to minimize the reflection coefficients and the mutual coupling between the feeding ports, and we apply a gradient-based optimization algorithm to solve the problem. 
The figure-of-merits are computed based on full-wave solutions of Maxwell's equations using our in-house finite-difference time-domain (FDTD) solver, and the required sensitivities are computed using the adjoint-field method.
We present the result of two optimized designs exhibiting low reflection, high isolation, and enhanced bandwidth, demonstrating the effectiveness of the proposed topology optimization approach for the design of compact DP microstrip antennas with enhanced bandwidth.


\section{Problem setup}
\begin{figure}[!htb]
\centering
\includegraphics[scale=0.95]{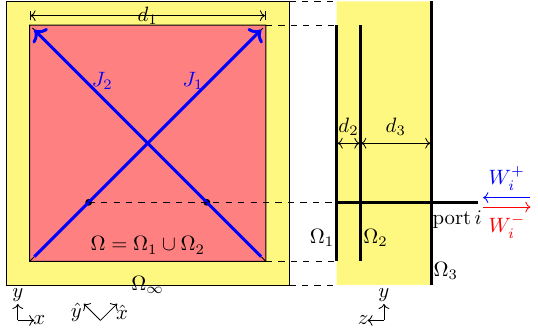}
\caption{Geometrical parameters of the two-layer microstrip antenna.
Top and side views illustrating the PCB stack-up used to implement the design. 
The square design domains $\Omega_1$ and $\Omega_2$, each with a side length~$d_1=26.3$\,mm, are separated by a distance~$d_2=0.21$\,mm and positioned~$d_3=1.28$\,mm above the ground plane $\Omega_3$. The two coaxial feed ports are connected at positions corresponding to one quarter of the diagonals. 
}
    \label{fig:setup}
\end{figure}

Fig.\,\ref{fig:setup} illustrates the problem setup, where a 4-layer FR-4 PCB stack-up is utilized. 
The square domains $\Omega_1$ and $\Omega_2$ represent the design regions in which the radiating elements of a two-layer microstrip antenna are to be designed. 
The antenna is fed through the ground plane $\Omega_3$ using two coaxial ports placed along the diagonals of the design domains, each located a distance of one-quarter of the diagonal from the center.
The diagonals are aligned with the $\hat{x}$ and $\hat{y}$ directions, as illustrated in Fig.\,\ref{fig:setup}(a). 
To enforce dual polarization, the design domains are constrained to be left–right, up–down, and diagonally symmetric. 
Placing the feeding ports along the diagonal $\hat{x}$ and $\hat{y}$ can improve the port isolation, since it results in a larger port separation distance compared to using ${x}$ and ${y}$.
By the imposed symmetry, when the antenna is excited through port\,1 (port\,2), the structure is expected to support the surface current $J_1$ ($J_2$), as illustrated in Fig.\,\ref{fig:setup}(a). 
With isolated ports, the orthogonal currents $J_1$ and $J_2$ give rise to dual-polarized radiation in the far-field. 
Furthermore, by optimizing the copper distribution in the design domain $\Omega=\Omega_1 \cup \Omega_2$, we aim at achieving a low reflection coefficient and strong isolation between the two feeding ports.







\section{Optimization formulation}
The time-domain Maxwell's equations are solved inside the analysis domain described in Section\,\ref{sec:Results}. 
The ports are fed by coaxial cables supporting the TEM mode, and the outgoing energy at port $i$ can be evaluated~\cite{hassan2014topology} using integral
\begin{align}\label{EnergyObjectve}
&W_{i}^-=\frac{1}{4Z_c}\int_0^T(V-Z_cI)^2\,dt,
\end{align}
where $V$, $I$, and $Z_c$ are the voltage, the current, and the characteristic impedance of the feeding cable, and $T$ is the total simulation time. 
The antenna system satisfies the following energy balance,
\begin{equation}
\sum_{i=1}^2W_i^++W_{\text{in},\Omega_\infty}=W_{\text{out},\infty}+\sum_{i=1}^2W_i^- +W_{\Omega_\infty},
\end{equation}
where $W_i^+$ and $W_{\text{in},\Omega_\infty}$ denote the supplied energies at port $i$ and from a far-field source, respectively; $W_{\text{out},\Omega_\infty}$ denotes the outgoing energy radiated by the system; and $W_{\Omega_\infty}$ denotes the energy loss in the system. 
We formulate the optimization problem
\begin{equation}
\begin{aligned}
 \min_{\sigma}{ \log{\left( \frac{\mathcal{D}_{1,1}\left(\sigma\right) \,\mathcal{D}_{2,1}\left(\sigma\right)}{ \mathcal{D}_{1,
p}\left(\sigma\right)}\right)}},
\end{aligned}
\label{eq:obj_fnopt}
\end{equation}
subject to the governing equations, where $\sigma\in [\sigma_\text{min},\sigma_\text{max}]$ denotes the conductivity in the design domains.
We use $\sigma_{\text{min}}=10^{-4}$\,S/m and $\sigma_\text{max}=10^5$\,S/m to represent the conductivities of a good dielectric and a good conductor, respectively.
The quantity $\mathcal{D}_{i,j}=(1+W_{i,j}^{-})^{q_{i,j}}$ denotes the weighted and regularized outgoing energy at port $i$ when port $j$ is excited, where the subscript $p$ denotes a plane wave source (that is, the antenna operates in its receiving mode\,\cite{hassan2014topology}). 
The parameters $q_i$ are used to control the relative scaling and emphasis of the sub-objectives in problem\,\eqref{eq:obj_fnopt}.

Including the maximization of the outgoing energy $\mathcal{D}_{1,p}$ in the objective function formulation enforces the design material to be less lossy, that is, to be a good dielectric or a good conductor. 
Without this term, the reflection coefficient $\mathcal{D}_{1,1}$ could be trivially minimized by employing lossy designs\,\cite{hassan2014topology}. 
The minimization of the term $\mathcal{D}_{2,1}$ maximizes the isolation between the two ports. 
Due to the symmetry of both the structure and the ports' placements, only the excitation of one port (here, port\,1) needs to be considered when evaluating the objective function.

To solve problem\,\eqref{eq:obj_fnopt} using gradient-based optimization methods, the objective function gradient is calculated via the adjoint field method using,
\begin{equation}
    \delta W_{\text{i}}^-(\sigma,\delta\sigma)=-\int_\Omega\int_0^T \mathbf{E}(T-t)\cdot \mathbf{E}^*(t)\,\delta\sigma\, dt\,  d\Omega,
    \label{eq:portenergy}
\end{equation}
with the symbol $\delta$ denotes the first-order variation, $\mathbf{E}$ is the electric field in the design domain, and $\mathbf{E}^*$ is an adjoint field obtained by solving an adjoint field problem. 
The adjoint field problem consists of one additional solution of Maxwell's equation in the analysis domain, where the recorded outgoing signals from the forward problem are reversed in time and used as sources to feed their corresponding ports\,\cite{hassan2014topology}.

In density-based topology optimization, the design variables are typically associated with the entries of a density vector~$\bm{p}$ such that $p_i\in [0,1]$, with  $0$ and $1$ denoting the absence and presence of the design material, respectively. 
So-called filtering techniques are commonly employed to prevent convergence to low-performing local optima or to impose feature size control that might be useful for imposing manufacturing constraints. 
Here we apply a nonlinear filter that reduces intermediate densities and controls the minimum feature size.
The filter consists of $K$ consecutive filter operators $\mathbf F^k,k=1,\dots, K ,$ acting on the design vector~$\bm{p}$~\cite{hassan2018topology},
\begin{equation}
    \tilde{\bm{p}}=\mathbf{F}^{K}\left(  \mathbf{F}^{K-1}\left(\dots\mathbf{F}^1\left(\bm{p}\right) \right) \right)
\end{equation}
The filtered design variables $\tilde{\bm{p}}$ are mapped to the physical conductivity used in the simulation through
\begin{equation}
    \bm{\sigma}=10^{9\tilde{\bm{p}}-4}.
\end{equation}
We use the FDTD method with the convolutional perfectly matched layer (CPML)~\cite{cpml} to solve the 3D Maxwell's equations. 
Based on Yee's scheme\,\cite{taflove2005computational}, the discretized form of \eqref{eq:portenergy} is given by~\cite{hassan2014topology}
\begin{equation}\label{eq:DW-dsigmae}
\frac{\partial W_i^-}{\partial \sigma_{e}}=-({h)}^3\sum_{n=1}^N E^{N-n}_e
\frac{E^{*n-\frac{1}{2}}_e+E^{*n+\frac{1}{2}}_e}{2}\Delta t,
\end{equation} 
where $\sigma_e$ is the conductivity at edge $e$ in the mesh, $h$ and $\Delta t$ are the spatial and temporal discretization steps, respectively, $E_{e}$ and $E^*_{e}$ are the electric and adjoint fields, respectively, and $N$ is the number of total simulation time steps.
From expression~\eqref{eq:DW-dsigmae} and the mapping $\bm p \mapsto\sigma_e$ described above, the gradient of the objective function with respect to the density variables is computed using the chain rule.
We use the globally convergent method of moving asymptotes (GCMMA)~\cite{SvanbergGlobally} to update the design vector and iteratively solve problem\,\eqref{eq:obj_fnopt}.

\section{Results}\label{sec:Results}
We use our in-house 3D\,FDTD solver to compute the solutions of both the forward and adjoint systems\,\cite{ludecouple}. 
The solver is implemented with the CUDA toolkit to run on graphics processing units (GPUs) and is executed on computing resources provided by HPC2N\,\cite{hpc2n}, using compute nodes equipped with AMD Zen4 CPUs and NVIDIA H100 GPUs. 
The time and spatial discretization steps used in the FDTD simulations are $\Delta t={0.99h}/{\sqrt{3}c}$ and $h=0.10$\,mm, respectively, with $c$ denoting the speed of light in vacuum. 
The total simulation domain including the CPML layers is discretized into $310\times310\times69$ Yee cells, and each layer of the design area consists of $250\times250$ cells, resulting in a total number of $251\,000$ design variables.
The number of time steps used in each FDTD simulation is $N = 30\,000$.  
Each optimization iteration requires two forward and two adjoint simulations: one simulation when port\,1 is excited (transmitting mode) and another simulation when the plane wave is used as a source (receiving mode).
As an excitation signal, we use a modulated \emph{sinc} pulse with a half-power bandwidth of 1.5\,GHz and centered around 5.5\,GHz.

Based on the abovementioned symmetric conditions, the design domains in Fig.\,\ref{fig:setup} is divided into eight symmetric parts to ensure dual polarization. 
The symmetry is enforced on both the design variables and the gradients computed for use in GCMMA. 
As the two layers are optimized simultaneously, we allow only one layer to be connected to the inner probe of the coaxial cable, while the other layer is kept disconnected.
This approach enables the disconnected two layers to develop different resonance modes.
Initially, the design variables are uniformly set to $\rho_i=0.7$, except for a $10 \times 10$\,cells region around the probe in the disconnected layer, where $\rho_i=0$.  
The weighting factors $q_{i,j}$ are determined based on numerical investigations to provide an appropriate balance between the different sub objectives.
The final optimized design is thresholded around $\rho_i=0.5$ to void (if $\rho_i<0.5$) or the conductivity of copper (if $\rho_i\geq0.5$). 

\begin{figure}[!htb]   
\centering
\includegraphics[width=0.95\columnwidth]{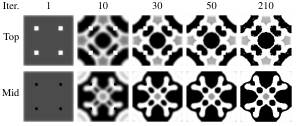}
\caption{Optimization history of Design\,I at different iterations.}  
\label{fig:DesignIHis}
\end{figure}
\begin{figure}[!htb]   
\centering
\includegraphics[height=3cm]{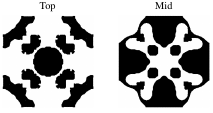}
\caption{Optimized two-layer microstrip antenna, Design\,I. }  
\label{fig:DesignI}
\end{figure}
\begin{figure}[!htb]   
\begin{subfigure}[t]{0.31\columnwidth}
\includegraphics[height=0.8\columnwidth]{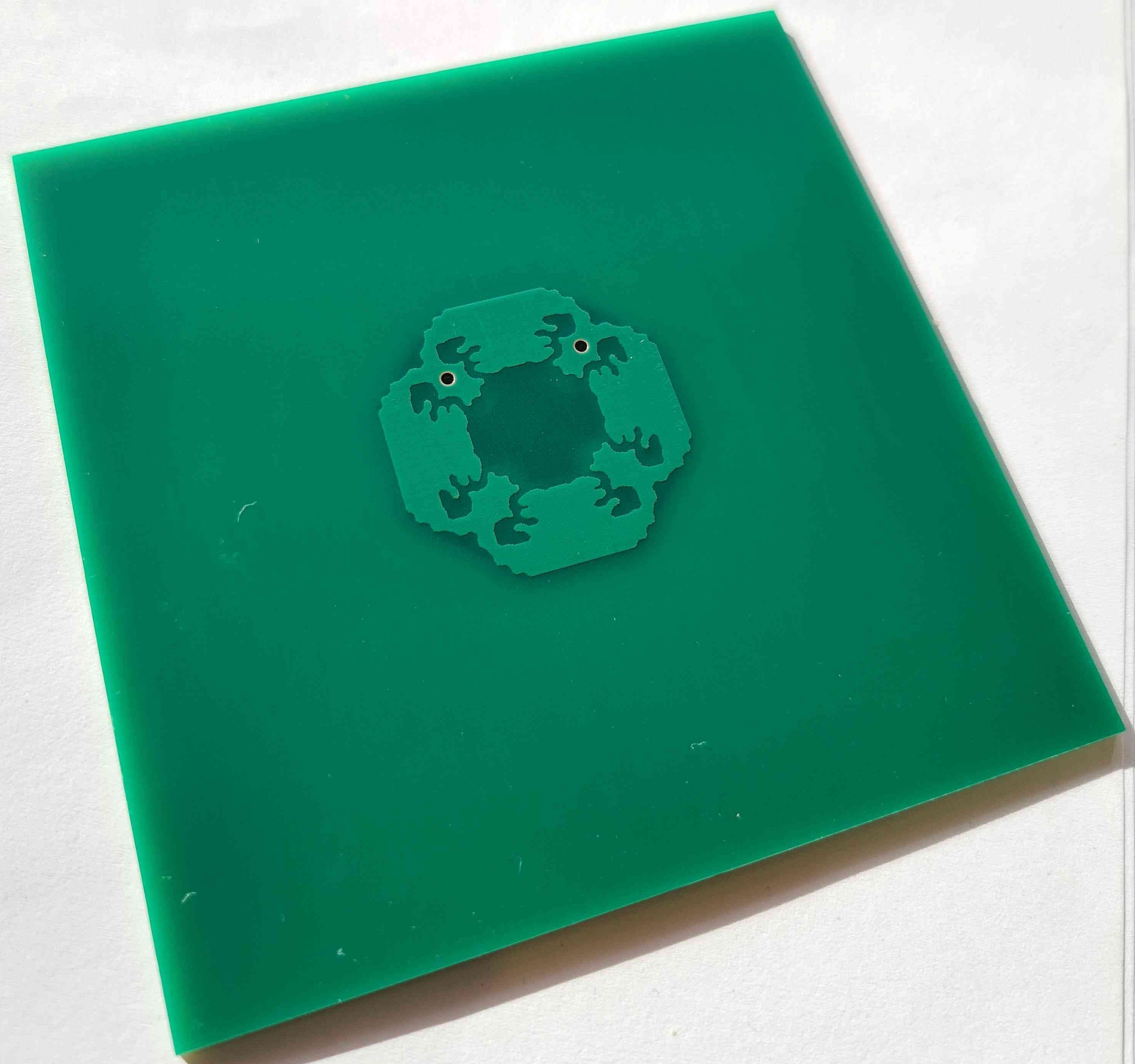}
\caption{}
\end{subfigure}  
\begin{subfigure}[t]{0.31\columnwidth}
\includegraphics[height=0.8\columnwidth]{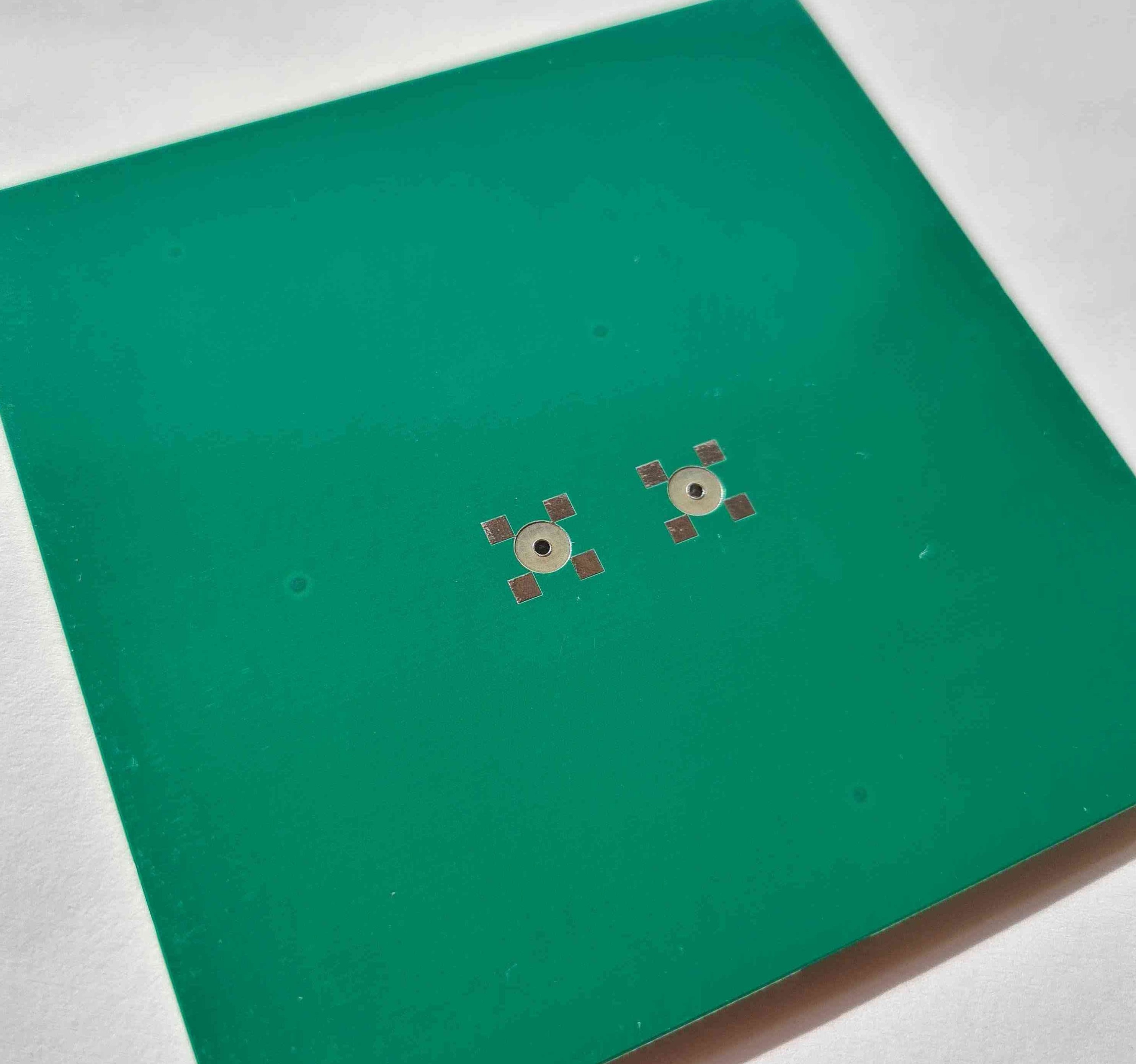}
\caption{}
\end{subfigure} 
\begin{subfigure}[t]{0.31\columnwidth}
\includegraphics[height=0.8\columnwidth]{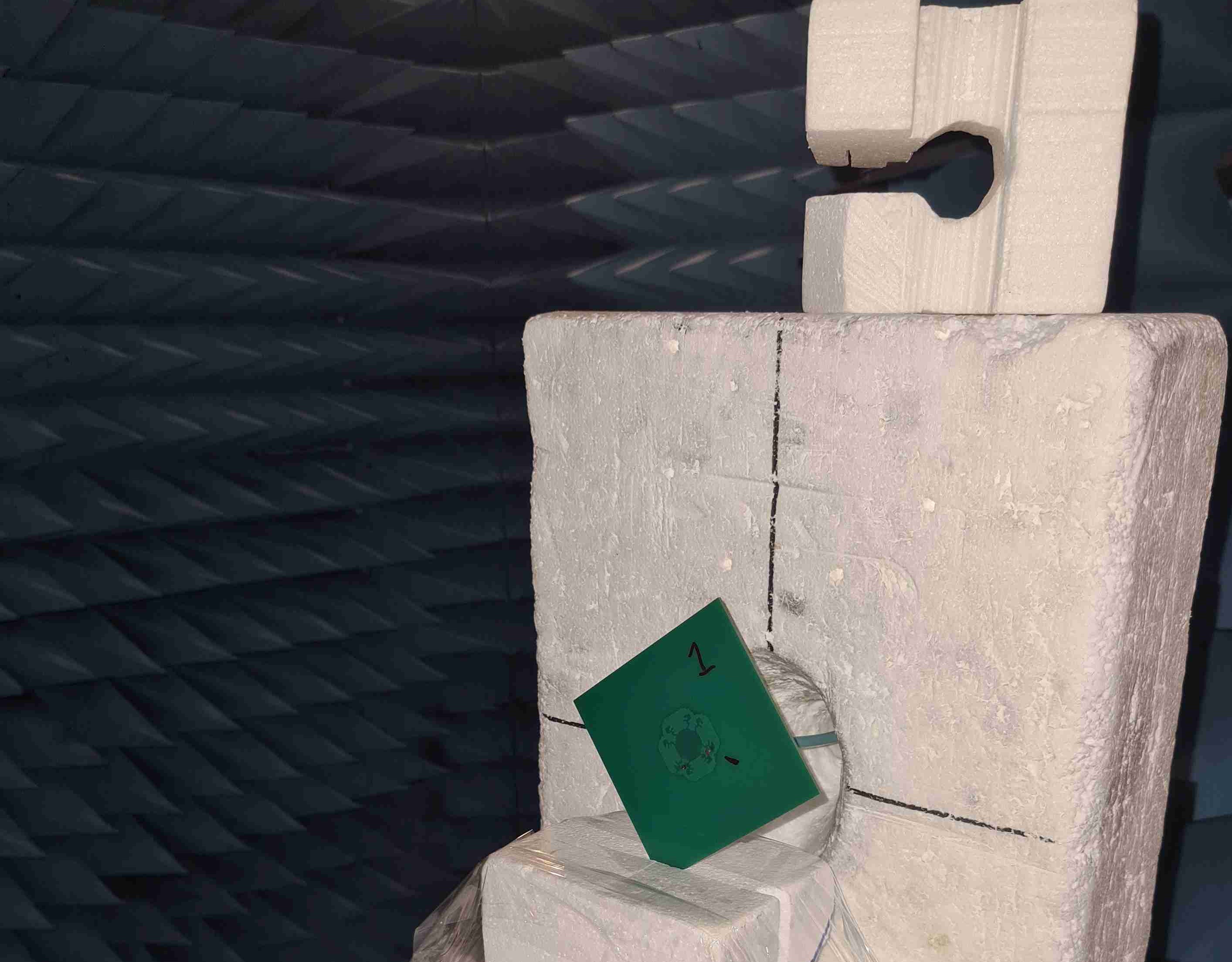}
\caption{}
\end{subfigure} 
\caption{Photographs of one design and the measurement setup: (a) top and (b) bottom views of the antenna, (c) measurement setup.}  \label{meassetup}
\end{figure}

We present the results of two optimized designs.
Fig.\,\ref{fig:DesignIHis} shows the optimization history of the first design, denoted as Design\,I, in which the middle layer is connected to the probe while the top layer is disconnected.
The optimization algorithm terminated after 210~iterations thresholded to the final design shown in Fig.\,\ref{fig:DesignI}.
Fig.~\ref{meassetup} shows one of the fabricated designs and the measurement setup.

\begin{figure}[!htb]   
\begin{subfigure}[t]{0.24\textwidth}
\centering
\includegraphics[width=0.8\columnwidth,  trim=0cm 0.0cm 0cm 0cm, clip]{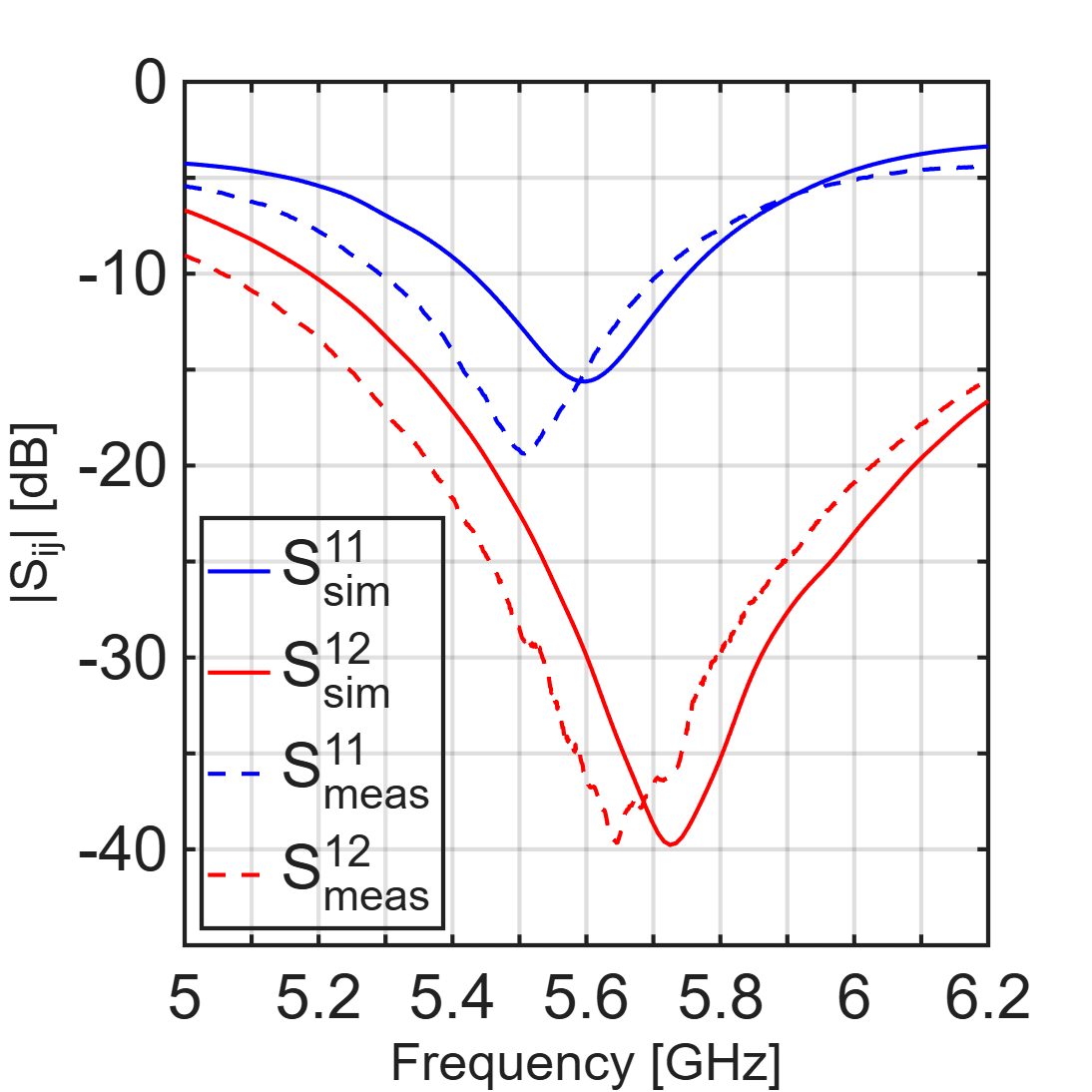}
\caption{}\label{fig:DesignISpara}
\end{subfigure}  
\begin{subfigure}[t]{0.24\textwidth}
\centering
\includegraphics[width=0.8\columnwidth,  trim=0cm 0.0cm 0cm 0cm, clip]{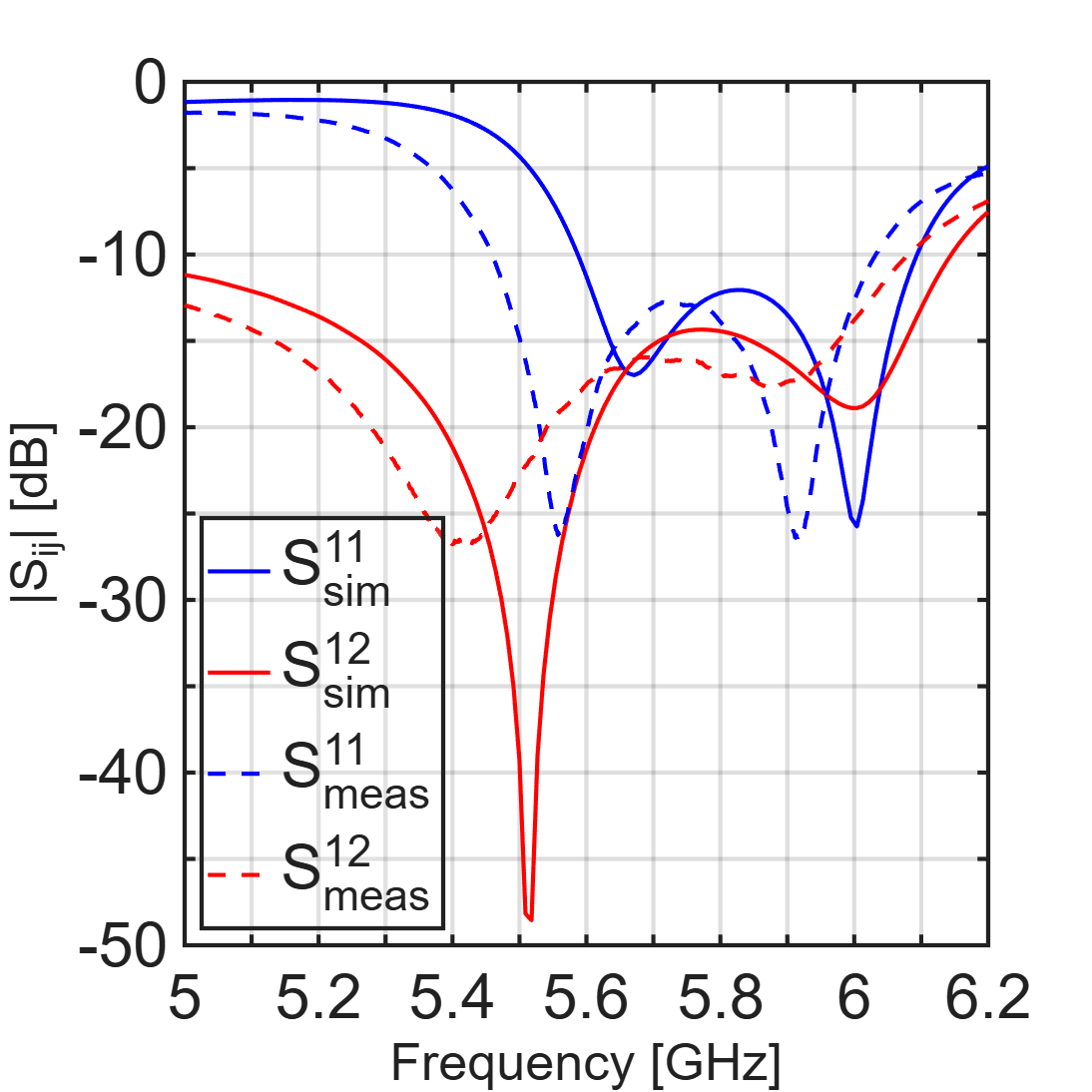}
\caption{}\label{fig:DesignIISpara}
\end{subfigure}  
\caption{Simulated and measured S–parameters of (a) Design\,I and (b) Design\,II.}  
\end{figure}

Fig.~\ref{fig:DesignISpara} shows the simulated and measured S-parameters of Design\,I. 
In simulation, the antenna resonates at 5.59\,GHz with $|S_{11}| \approx -14$\,dB. 
The $-10$\,dB impedance bandwidth spans the frequency range 5.43\,GHz to 5.75\,GHz, corresponding to a fractional bandwidth of approximately 5.8\%. 
Due to fabrication tolerances, the measured $-10$\,dB bandwidth shifts toward lower frequencies, spanning from 5.30\,GHz to 5.71\,GHz, with a slightly increased fractional bandwidth of about 7.5\%.
Although the $-10$\,dB impedance bandwidth is moderate, the $|S_{11}|$ remains below $-5$\,dB over a wider frequency range.
The magnitude of the mutual coupling $|S_{21}|$ between the two ports remains lower than $-20$~dB over most of the operating frequency band, which confirms the high isolation between the two ports.

\begin{figure}[!htb]   
\centering
    \includegraphics[height=3cm]{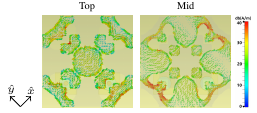}
\caption{Current distribution of Design\,I at 5.7\,GHz (port\,1 is excited). }  
\label{fig:DesignI_sd}
\end{figure}

The simulated current distribution is shown in Fig.~\ref{fig:DesignI_sd}. 
The current distributions of the two metallic layers when port\,1 is excited are nearly symmetric around the diagonal $\hat{x}$; and the influence of the feeding probes is not significant, confirming that the symmetry conditions successfully enforce a well-defined orthogonal polarization mode.

\begin{figure}[!htb]   
\begin{subfigure}[t]{0.24\textwidth}
\centering
\includegraphics[width=0.95\columnwidth,  trim=0cm 1.6cm 0cm 0cm, clip]{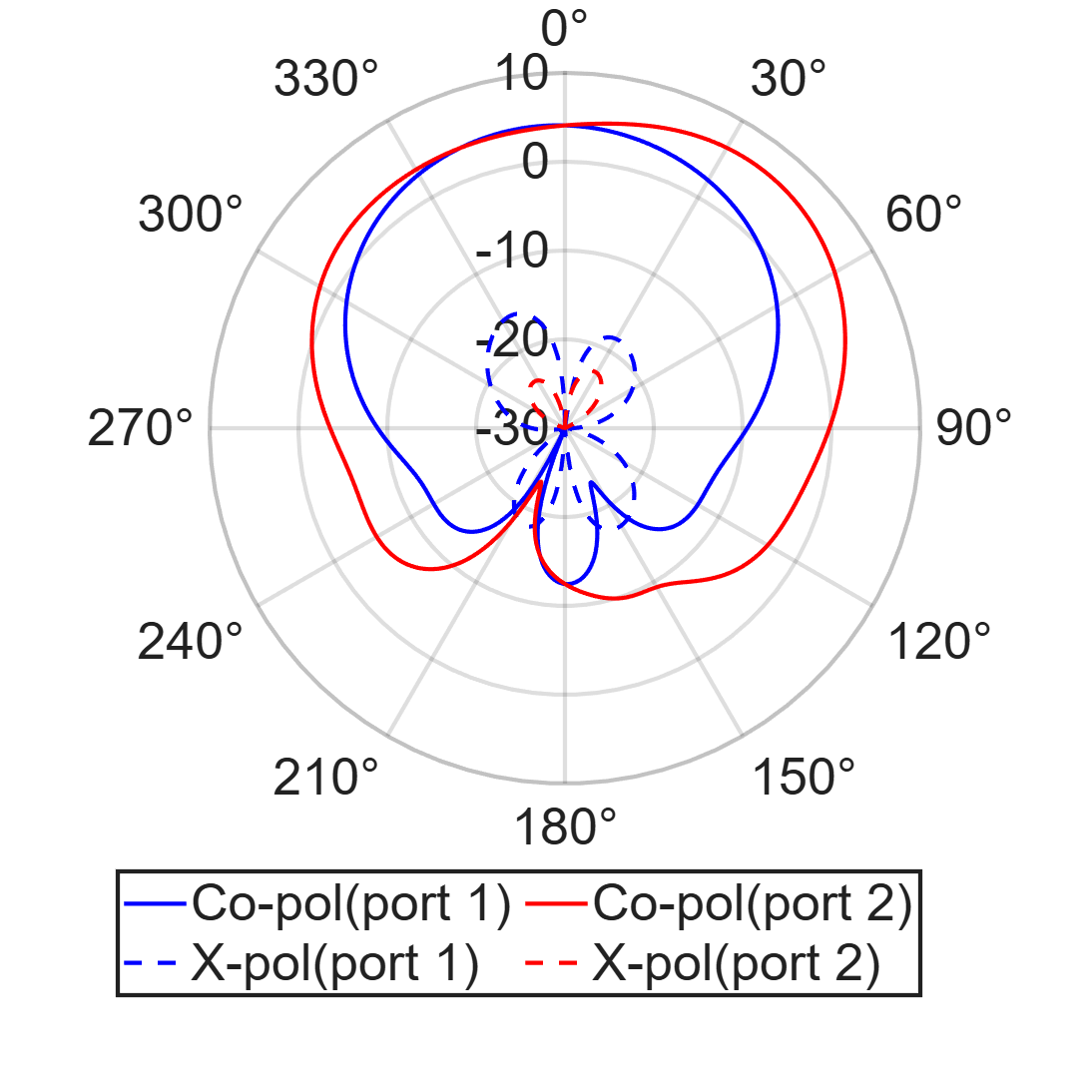}
\caption{}
\end{subfigure}  
\begin{subfigure}[t]{0.24\textwidth}
\centering
\includegraphics[width=0.95\linewidth,  trim=0cm 1.6cm 0cm 0cm, clip]{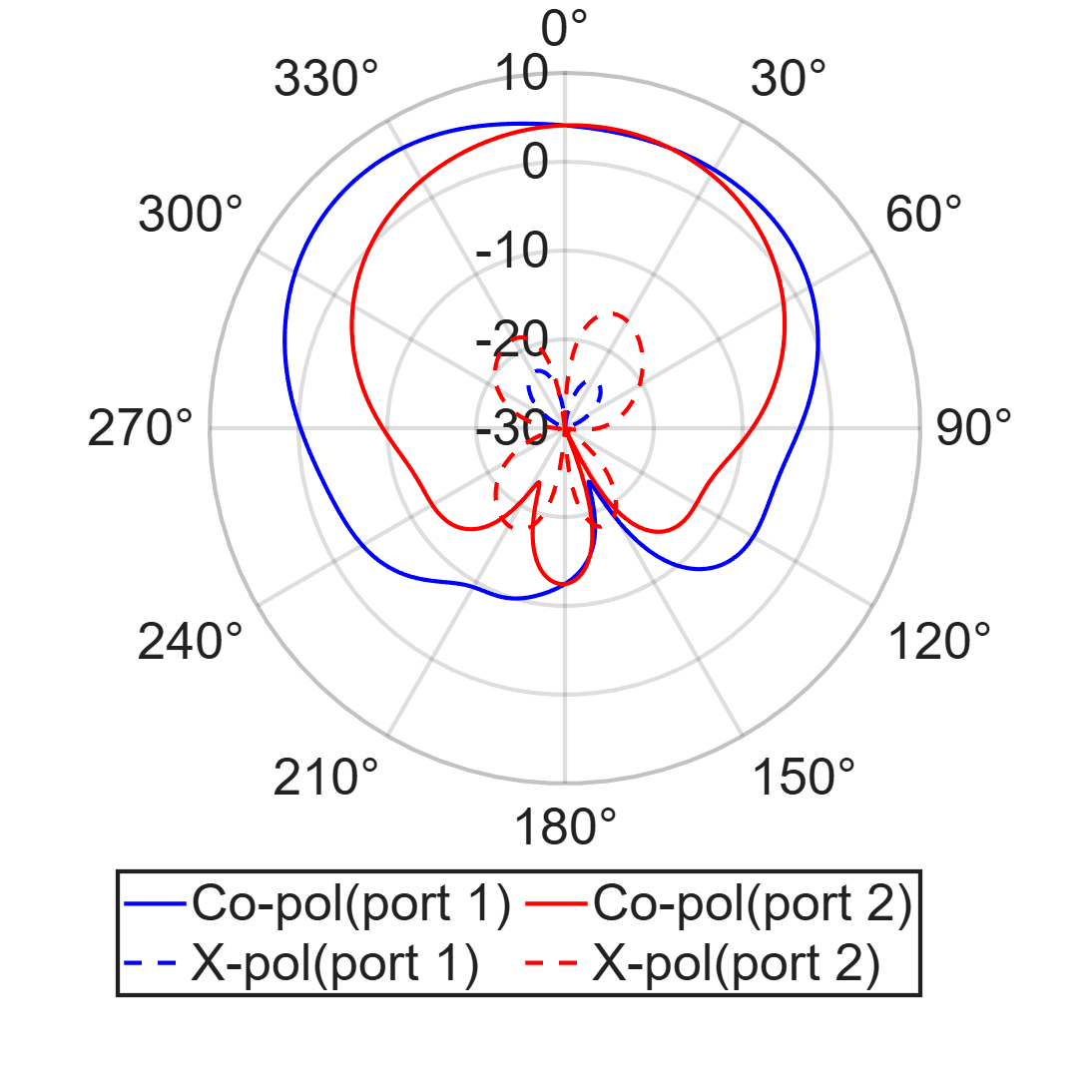}
\caption{}
\end{subfigure}

\begin{subfigure}[t]{0.24\textwidth}
\centering
\includegraphics[width=0.95\linewidth,  trim=0cm 1.6cm 0cm 0cm, clip]{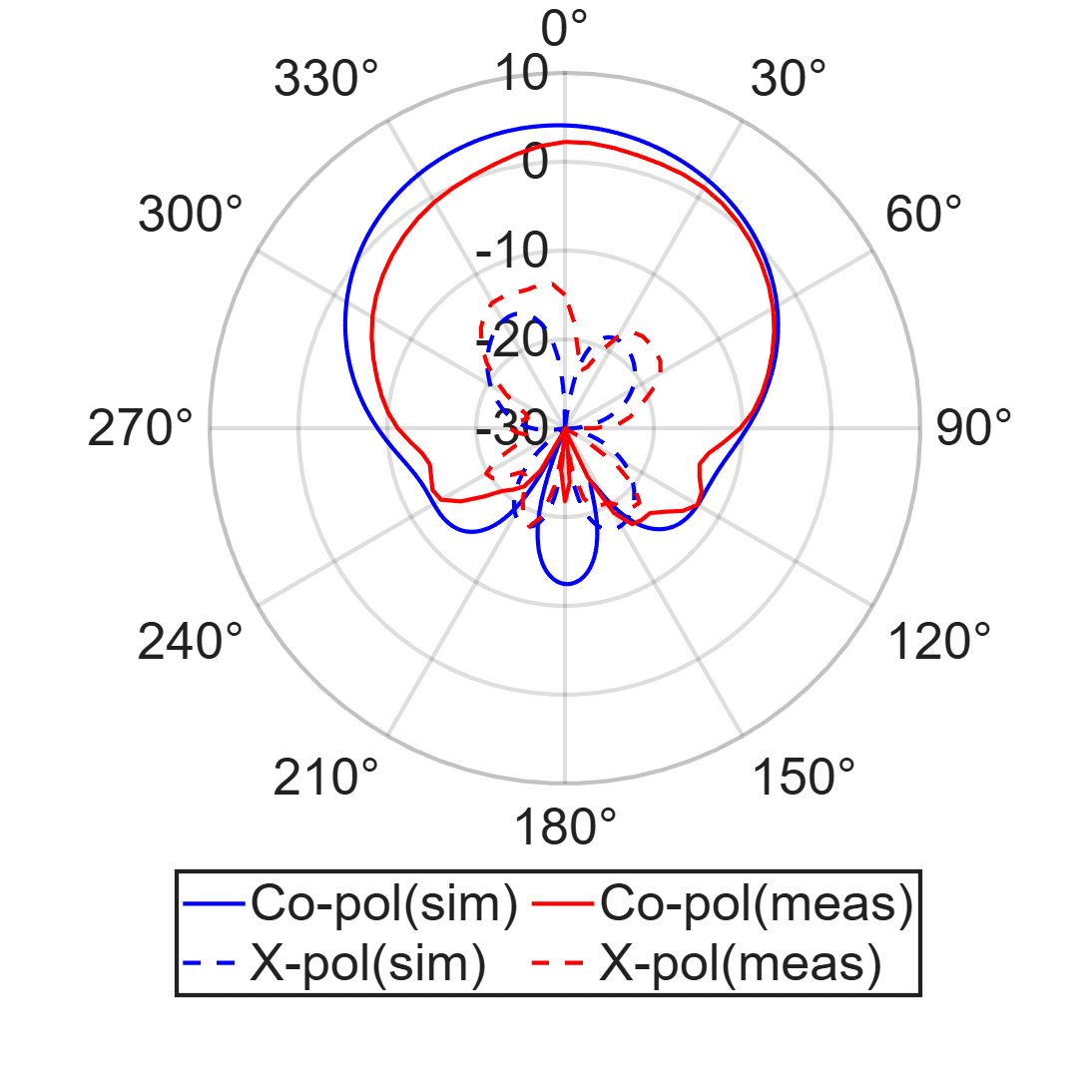}
\caption{}\label{fig:DesignI_com1}
\end{subfigure}
\begin{subfigure}[t]{0.24\textwidth}
\centering
\includegraphics[width=0.95\linewidth,  trim=0cm 1.6cm 0cm 0cm, clip]{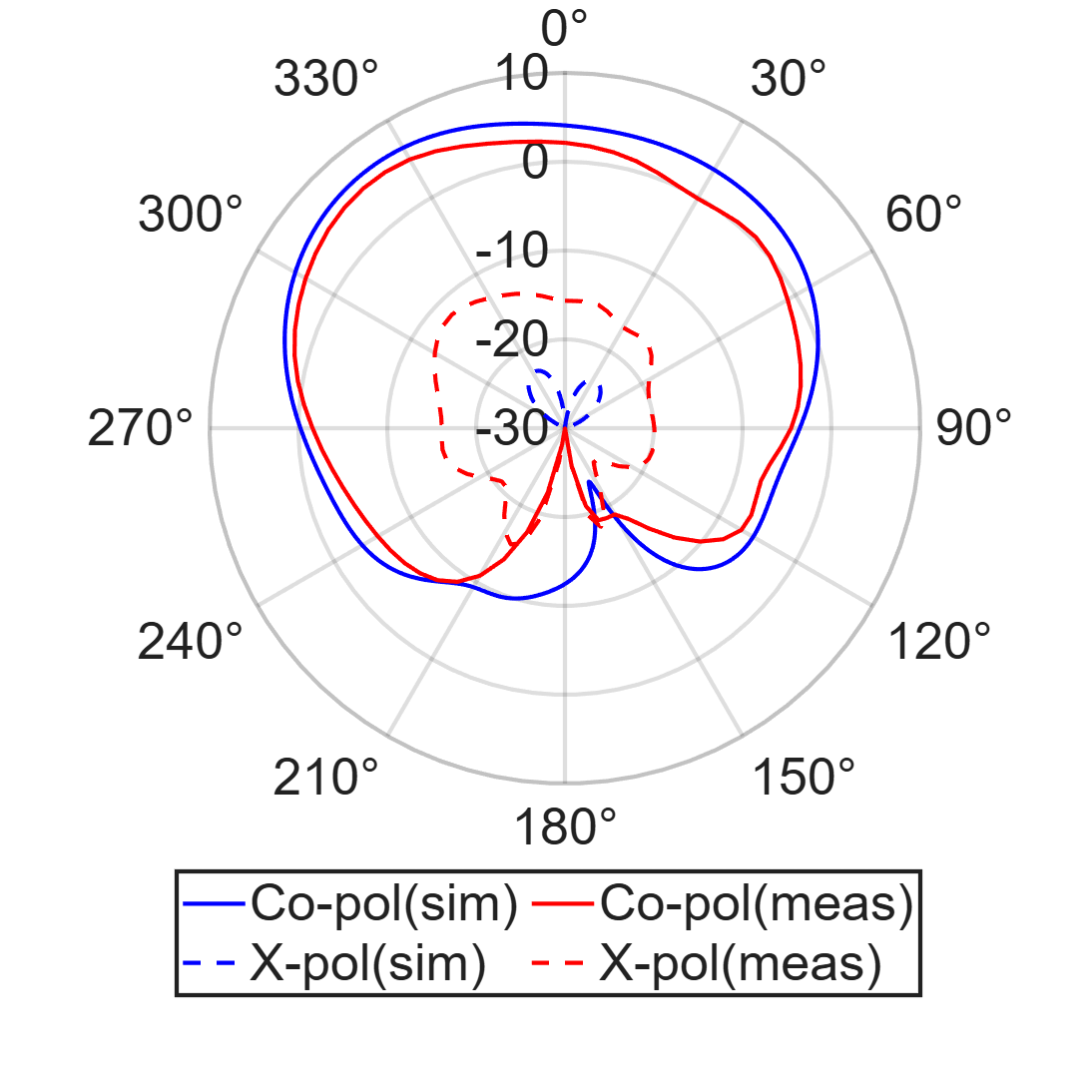}
\caption{}\label{fig:DesignI_com2}
\end{subfigure}
\caption{Radiation patterns of Design\,I, where Co-pol and X-pol represent the co-polarization and cross-polarization, respectively. Simulation: (a) $\hat{x}oz$ plane, (b) $\hat{y}oz$ plane. Comparison between simulation and measurement when port\,1 is excited: (a) $\hat{x}oz$ plane, (b) $\hat{y}oz$ plane.}  
\label{fig:DesignI_sim}
\end{figure}

The radiation patterns of the DP microstrip antenna at 5.7 GHz are shown in Fig.~\ref{fig:DesignI_sim}, including a comparison between the simulation and measurement when port\,1 is excited. 
The maximum realized gain of the DP microstrip antenna is 6.61 dBi. The radiation patterns show that the $\hat{x}oz$ plane for port\,1 excitation and the $\hat{y}oz$ plane for port\,2 excitation are nearly symmetric. As shown in Figs~\ref{fig:DesignI_com1} and~\ref{fig:DesignI_com2},  the simulated and measured patterns match well with each other.



\begin{figure}[!htb]   
\centering
\includegraphics[height=3cm]{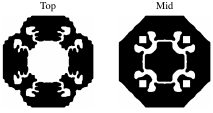}
\caption{Optimized two-layer microstrip antenna, Design\,II.}  
\label{fig:DesignII}
\end{figure}

\begin{figure}[!htb]   
\centering
    \includegraphics[height=3cm]{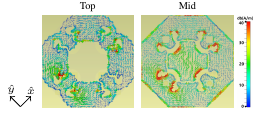}
\caption{Current distribution of Design\,II at 5.7\,GHz (port\,1 is excited). }  
\label{fig:DesignII_sd}
\end{figure}

\begin{figure}[!htb]   
\begin{subfigure}[t]{0.24\textwidth}
\centering
\includegraphics[width=0.95\columnwidth,  trim=0cm 1.6cm 0cm 0cm, clip]{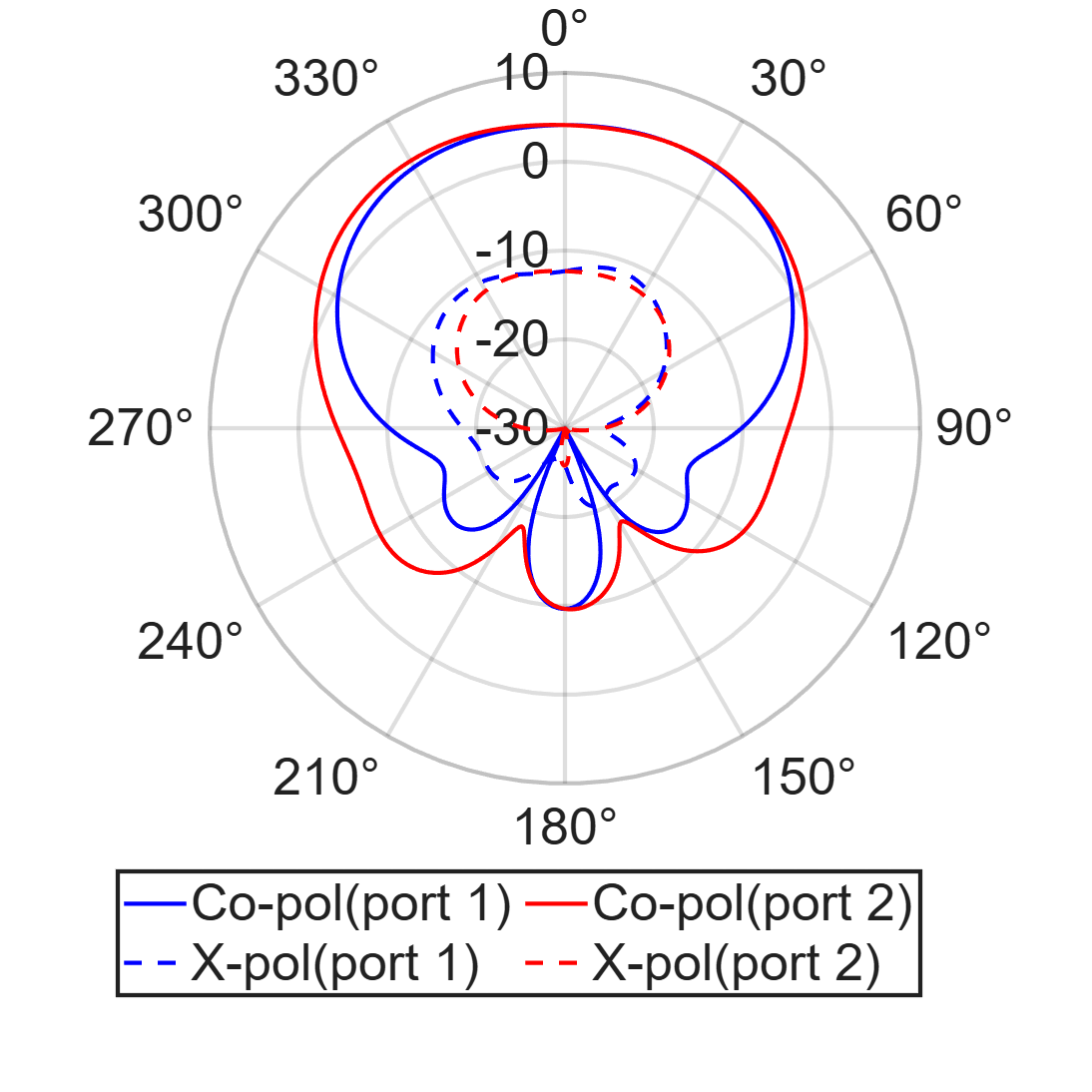}
\caption{}\label{fig:DesignII_far1}
\end{subfigure}  
\begin{subfigure}[t]{0.24\textwidth}
\centering
\includegraphics[width=0.95\linewidth,  trim=0cm 1.6cm 0cm 0cm, clip]{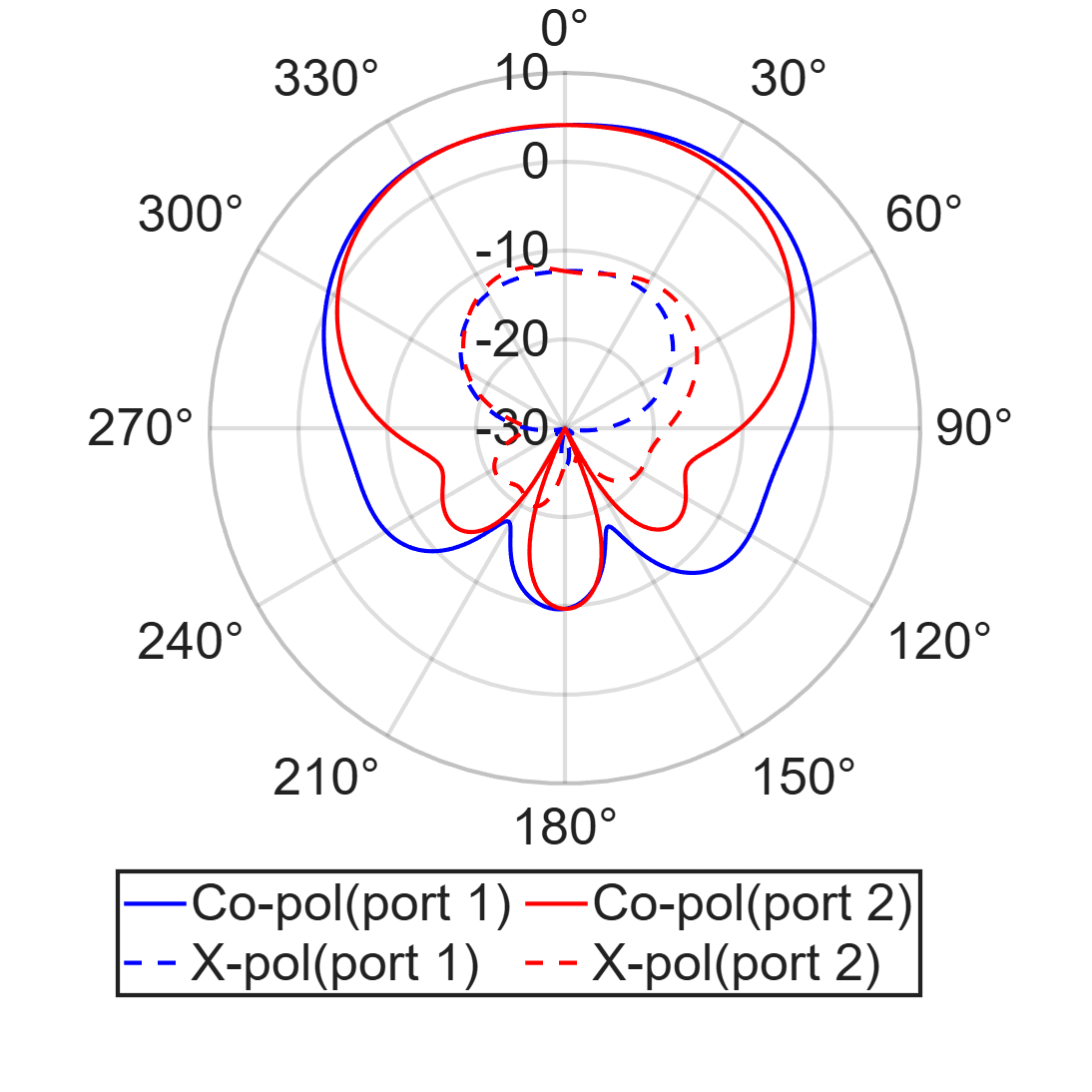}
\caption{}\label{fig:DesignII_far2}
\end{subfigure}
\begin{subfigure}[t]{0.24\textwidth}
\centering
\includegraphics[width=0.95\columnwidth,  trim=0cm 1.6cm 0cm 0cm, clip]{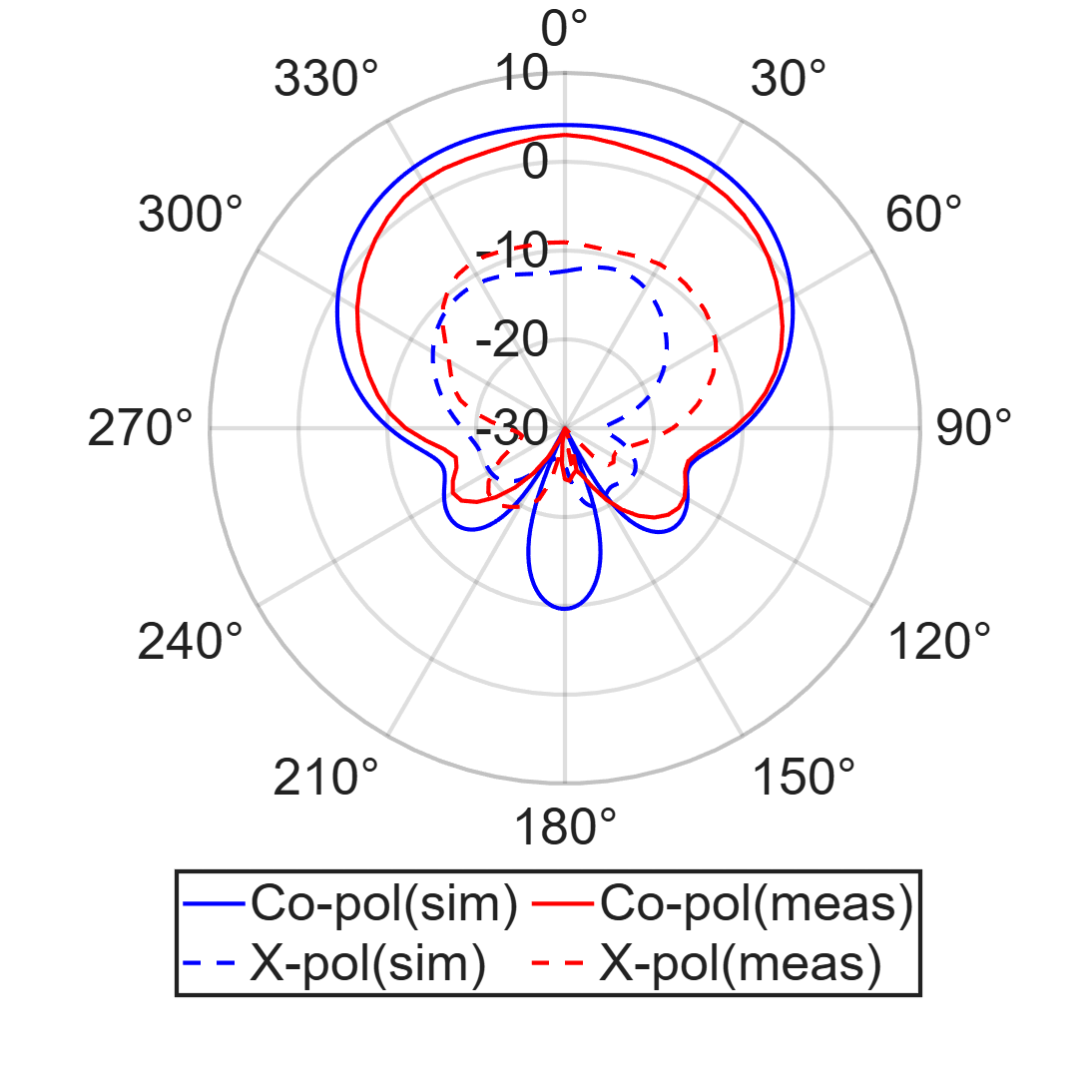}
\caption{}
\end{subfigure}  
\begin{subfigure}[t]{0.24\textwidth}
\centering
\includegraphics[width=0.95\linewidth,  trim=0cm 1.6cm 0cm 0cm, clip]{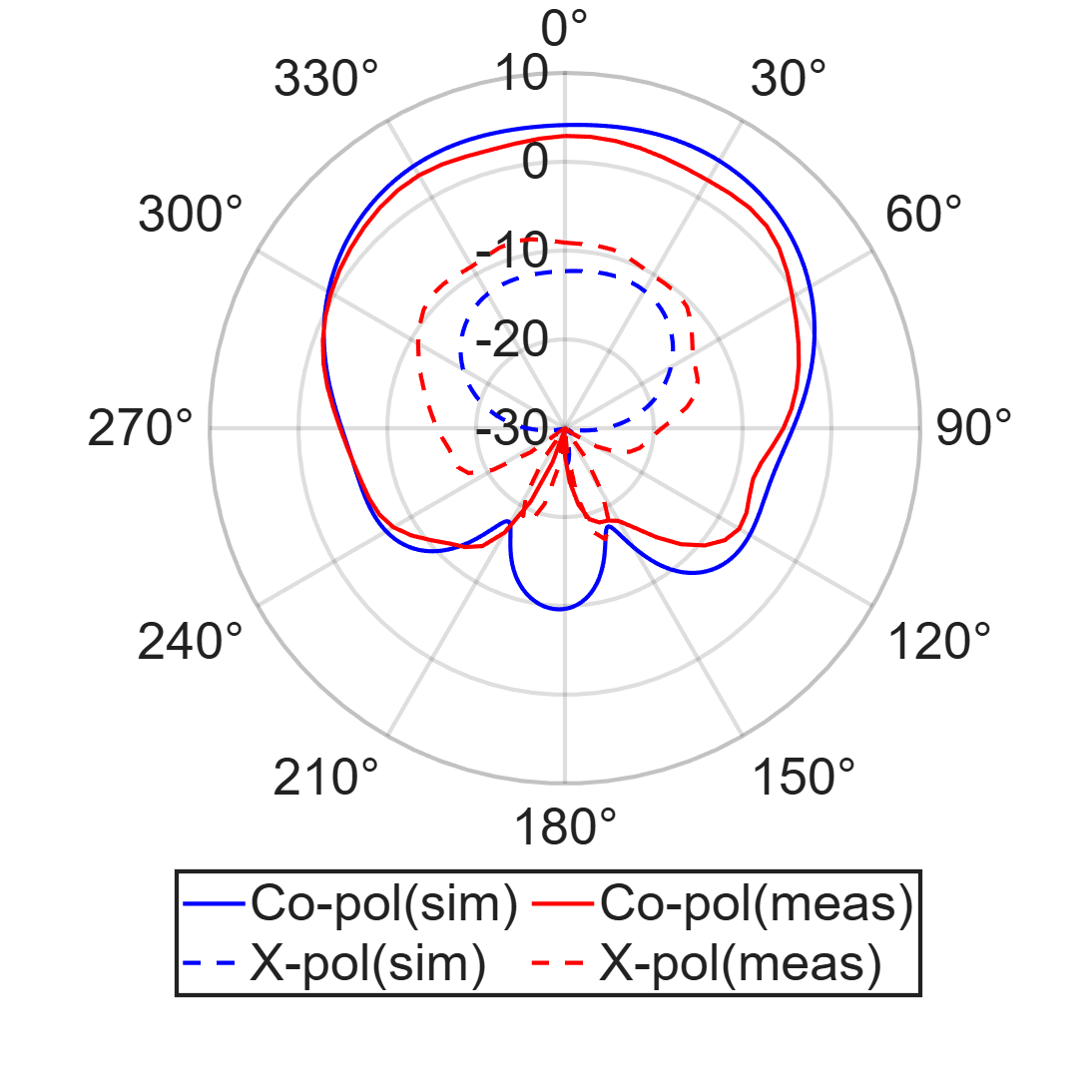}
\caption{}
\end{subfigure}
\caption{Radiation patterns of Design\,II, where Co-pol and X-pol represent the co-polarization and cross-polarization, respectively. Simulation: (a) $\hat{x}oz$ plane, (b) $\hat{y}oz$ plane. Comparison between simulation and measurement when port\,1 is excited: (c) $\hat{x}oz$ plane, (d) $\hat{y}oz$ plane.}  
\label{fig:DesignII_sim}
\end{figure}

Fig.~\ref{fig:DesignII} shows the optimization results for the second case study, where the probe is connected to the top layer. 
In this case, denoted as Design\,II, the top layer acts as the active radiating element, while the middle layer serves as a parasitic element supporting the design objectives. 
Fig.~\ref{fig:DesignIISpara} shows the simulated and measured S-parameters of Design\,II. 
Two resonances are observed at 5.67\,GHz and 6.00\,GHz in simulation and the $-10$ dB bandwidth extends to a wider range from 5.59\,GHz to 6.09\,GHz. The measured $|S_{11}|$ have two peak resonances at 5.56\,GHz and 5.92\,GHz with minimum value around $-26$ dB, with the $-10$ dB bandwidth from 5.46\,GHz to 6.03\,GHz  corresponding to a fractional bandwidth of 9.9\%.
Moreover, the isolation between the two ports remains more than 15\,dB across the same frequency range.
Fig.~\ref{fig:DesignII_sd} shows the simulated current distribution of the two-layer antenna when port\,1 is excited.
The current distribution is consistent with the symmetry imposed on the design.
The radiation patterns of Design~II at 5.7\,GHz are shown in Fig.~\ref{fig:DesignII_sim}. 
The simulated patterns confirm that the two ports excite symmetric and orthogonal radiation modes. The measured results exhibit similar radiation characteristics as in the simulations with slight discrepancies.


Table\,\ref{tab:comparison} presents a comparison between Design\,II and some recent work on dual-polarized microstrip antennas.
It is challenging to fairly compare with other works due to differences in key design aspects, such as operating frequencies, substrate parameters, and design approaches.
Nevertheless, our design appears favorable with respect to its wideband performance.

\begin{table}
\centering\setlength{\tabcolsep}{1pt}
\caption{Comparison with other microstrip antennas.}
    \label{tab:comparison}
\begin{tabular}{|c|c|c|c|c|c|}
 \hline
\multirow{1}{*}{ Ref.}& \multirow{1}{*}{Fre.} & \multirow{1}{*}{Method}& {Layer}&{Size($\lambda_0$)}&\multirow{1}{*}{BW(\%)}\\
\hline
\multirow{1}{*}{\cite{chen2023single}}&{[4.74,4.83]}&TA&1&$1.44\times1.44\times0.024$&1.8\%\\\hline
\multirow{1}{*}{\cite{zhu2022hybrid}}&{[5.71,5.89]}&OA&1&$0.35\times0.35\times0.058$&3.1\%\\\hline
\multirow{1}{*}{\cite{he2020dual}}&{[2.48,2.56]}&TA&1&$0.19\times0.19\times0.07$&3.0\%\\\hline
\multirow{1}{*}{\cite{10824914}}&{[5.72,5.90]}&TA&1&$0.23\times0.23\times0.07$&3.1\%\\\hline
\multirow{1}{*}{This}&{[5.46,6.03]}&OA&2&$0.50\times0.50\times0.031$&9.9\%\\\hline
\multicolumn{6}{l}{Ref.:references; Fre.: frequency range (GHz); BW:bandwidth.}\\
\multicolumn{6}{l}{Method: theoretical analysis(TA), optimization algorithms\,(OA);}\\
\end{tabular}
\end{table}

\section{Conclusion}
This work proposes the design of a two-layer DP microstrip antenna with enhanced bandwidth using a topology optimization approach. 
The design problem is formulated to account for the near-field port matching and isolation, as well as the far-field dual polarizations.
The imposed symmetry conditions required to enforce the dual polarization restrict the active number of design variables for the optimization problem. 
However, using two copper layers for the antenna, from a standard FR4 stack-up, provides additional degrees of freedom, which enable the fulfillment of the multiple design requirements.
Our investigations suggest that a good performance can be achieved when the excitation ports are connected to only one layer, while the other layer serves as a parasitic element supporting the design objectives.
In particular, better results, particularly with respect to broadband response, are obtained when the outer design layer is actively connected to the probes.
Our results demonstrate the effectiveness of algorithm-based design as a supportive tool for employing multilayer PCBs to design compact, multiobjective microstrip antennas. 

\section*{Acknowledgments}
Partial funding for this work is provided by the Swedish strategic research program eSSENCE and Kempestiftelserna (Grant No. JCSMK23-0187).
The computations were performed on resources provided by the High Performance Computing Center North (HPC2N). 

\bibliographystyle{IEEEtran}
\bibliography{ref}

\vfill

\end{document}